\documentstyle[aps,prc,epsfig]{revtex}
\def\ba{\begin{eqnarray}}
\def\ea{  \end{eqnarray}}
\def\be{\begin{equation}}
\def\ee{  \end{equation}}
\def\bfg{\begin{figure}}
\def\efg{  \end{figure}}
\def\bce{\begin{center}}
\def\ece{  \end{center}}
\def\la{\langle}
\def\ra{\rangle}

\def\calH{{\cal H}}
\def\calO{{\cal O}}
\def\calQ{{\cal Q}}

\def\calV{{\cal V}}
\def\mrhozt{(m_\rho^{(0)})^2}
\def\mrhozf{(m_\rho^{(0)})^4}
\def\ch{{\rm ch}}
\def\sh{{\rm sh}}
\def\em{{\rm em}}
\def\ueta{{\underline\eta}}
\def\e{{\rm e}}
\def\ie{{\it i.e.}}
\def\eg{{\it e.g.}}
\def\etal{{\it et al.}}

\def\roughly#1{\mathrel{\raise.3ex\hbox{$#1$\kern-.75em%
\lower1ex\hbox{$\sim$}}}}

\def\gsim{\roughly>}

\def\Tr{{\rm Tr}\,}

\begin{document}
\twocolumn[\hsize\textwidth\columnwidth\hsize\csname @twocolumnfalse\endcsname

\title{Isospin Fluctuations in QCD and Relativistic Heavy-Ion Collisions}

\author{
Madappa Prakash$^{a}$\thanks{E-mail: prakash@snare.physics.sunysb.edu}\,,
Ralf Rapp$^{a}$\thanks{E-mail: rapp@tonic.physics.sunysb.edu}\,,
Jochen Wambach$^{b}$\thanks{E-mail: wambach@physik.tu-darmstadt.de} and
Ismail Zahed$^{a}$\thanks{E-mail: zahed@zahed.physics.sunysb.edu}}

\address{
$^a${\it Department of Physics and Astronomy, State University of New York, 
     Stony Brook, NY 11794}\\
$^b${\it Institut f\"ur Kernphysik, TU Darmstadt, Schlo{\ss}gartenstr.~9, 
          D-64289 Darmstadt, Germany}
 }

\date{\today}
\maketitle
\begin{abstract}

We address the role of fluctuations in strongly interacting matter
during the dense stages of a heavy-ion collision through its
electromagnetic emission.  Fluctuations of isospin charge are considered
in a thermal system at rest as well as in a moving hadronic fluid at
fixed proper time within a finite bin of pseudo-rapidity.  In the
former case, we use general thermodynamic relations to establish a
connection between fluctuations and the space-like screening limit of
the retarded photon self-energy, which directly relates to the
emissivities of dileptons and photons.  Effects of hadronic
interactions are highlighted through two illustrative calculations.
In the latter case, we show that a finite time scale $\tau$ inherent
in the evolution of a heavy-ion collision implies that equilibrium
fluctuations involve both space-like and time-like components of the
photon self-energy in the system.  Our study of non-thermal effects,
explored here through a stochastic treatment, shows that an early and
large fluctuation in isospin survives only if it is accompanied by a
large temperature fluctuation at freeze-out, an unlikely scenario in
hadronic phases with large heat capacity.  
We point out
prospects for the future which include: (1) A determination of the
Debye mass of the system at the dilute freeze-out stage of a heavy-ion
collision, and (2) A delineation of the role of charge fluctuations
during the dense stages of the collision through a study of
electromagnetic emissivities.


\end{abstract}
\vspace{0.58cm}
]


\section{INTRODUCTION}  

Charge fluctuations extracted through event-by-event measurements in
high-energy heavy-ion collisions have recently attracted attention as
a potential signature of Quark-Gluon Plasma (QGP)
formation~\cite{data}.  In the confined phase charges are carried by
hadrons in integer units of $e$, while in the deconfined phase they
are attached to quarks in fractional units.  This leads to charge
fluctuations and associated thermodynamic susceptibilities that are
distinctly different in the two phases~\cite{McL87}.  If significant 
primordial fluctuations survive the evolution of the collision, one
might hope to deduce the properties of deconfined matter at the early stages,
analogous to the manner in which primordial fluctuations in the early
universe are inferred through the detection of anisotropy in the
cosmic background radiation. Beginning with the expected signals in
rapidity distributions~\cite{early_work}, many sources and signals of
fluctuations in hadronic observables have been considered
(cf. \cite{Hei01} and references therein).  For more recent
discussions, see, for example, Refs.~\cite{AHM00,JK00,FW00,SS01}.

In experiments at the CERN SPS and RHIC to date, observed fluctuations
in hadronic multiplicities, kaon to pion ratios, and transverse
momenta agree reasonably well with model calculations of these
quantities at the dilute freeze-out stage of the system~\cite{Hei01}.
Such studies, however, have not unraveled the physics associated with
the QGP phase transition.  Whether future collider experiments at RHIC
and LHC will reveal tell-tale signatures in fluctuations of hadronic
observables alone remains to be seen.

It must be noted that the measured fluctuations in hadronic
observables result from a combination of several factors that include:
(1) The evolution of the created system from a dense partonic stage to
a dilute hadronic stage.  The duration of each stage is especially
critical inasmuch as the memory of the partonic stage is more likely
to be retained by the hadrons provided the hadronic stage is of
relatively short duration; and (2) The extent to which interactions
are different in the two phases.  A caveat, of course, is that if
duality between the partonic and hadronic worlds holds, equivalent
descriptions of observables may be achieved by considering
interactions in the two phases.  A good example is provided by the
cross sections ratio $R=\sigma_{e^+e^- \to hadrons}/\sigma_{e^+e^- \to
\mu^+\mu^-}$ for $\sqrt{s}\gsim 1.5$~GeV; in this case, interacting
hadronic states yield a description that is equivalent to one that
involves perturbative quarks and gluons.

Our principal objective in this paper is to emphasize that possibly
the only way to address the role of fluctuations during the dense
stages of the collision is to exploit electromagnetic probes in
combination with hadronic probes.  Towards this goal, we study
relations between electromagnetic radiation and thermal fluctuations
of isospin charge in a system at rest as well as in an expanding
hadronic fluid at fixed proper time and in 
appropriately chosen bins of pseudo-rapidity.

For a system at rest, we utilize general thermodynamic relations to
establish connections between fluctuations in matter and the
space-like screening limit of the retarded photon self-energy, a
quantity which directly controls the thermal emissivities of dileptons
and photons. We consider two examples that exemplify the effects of
hadronic interactions on electromagnetic radiation.  In addition, we
point out the close connections that exist between fluctuations,
susceptibilities, and average phase space densities. At the dilute
freeze-out stage, the average phase space densities of pions and kaons
can be inferred through interferometric (Hanbury-Brown and Twiss)
studies~\cite{Ber94,Fer99} and provide for a direct determination of
the system's Debye mass.

For a system undergoing relativistic expansion, we derive a general
formula for fluctuations in isospin charge binned in pseudo-rapidity
intervals of width $\Delta$.  The constituents of the moving fluid, be
they partons or hadrons, are assumed to be in local thermal
equilibrium. In the case of hadrons, we include finite pion and baryon
chemical potentials.  We cast our result in a readily usable form for
the analysis of experiments, which have unavoidable restrictions in
rapidity and transverse momenta due to limited acceptance in these
variables.  We show that finite time scales $\tau$ inherent in
heavy-ion collisions imply that equilibrium fluctuations involve both
space-like and time-like components of the photon's polarization
function with predominant contributions arising from frequency
$f\approx 2\pi/\tau$ and wavelength $\lambda \approx \pi\tau\,{\rm
sh}(\Delta/2)$.  We also consider effects of non-equilibrium (for
previous work, see for example, Ref.~\cite{TF}) with a view to explore
the consequences of possible temperature fluctuations at the
freeze-out stage.


This article is organized as follows.  In Sec.~\ref{sec_rest}, the
relationship between isospin fluctuations and space-like physics that
leads to screening of charges is established in an infinite volume
system at rest and in thermal equilibrium.  Secs.~\ref{sec_rest}.A
through \ref{sec_rest}.D briefly summarize known relations that have
enabled us to interrelate fluctuations, susceptibilities, and average
phase space densities. In Sec.~\ref{sec_rest}.E, connections to
electromagnetic (dilepton and photon) emissivities are set up.  Two
illustrative calculations incorporating effects of hadronic
interactions in the dilepton and photon emission rates are presented
in Sec.~\ref{sec_rest}.F.  In Sec.~\ref{sec_moving}, the role of
fluctuations in heavy-ion collisions, in which the system created
expands from an initially dense partonic system to a dilute hadronic
system prior to freeze-out, is discussed with particular emphasis on
electromagnetic probes. We demonstrate that in a relativistically
expanding fluid, thermal fluctuations probe both space-like
(screening) and time-like (emission) physics (Sec.~\ref{sec_moving}).
A discussion that relates our findings to those in earlier works is
contained in Sec.~\ref{others}.  In Sec.~\ref{sec_offeq}, we show how
isospin fluctuations present in the early stage can only survive if a
sizable temperature fluctuation is simultaneously present at
freeze-out.  This section contains an analysis of non-equilibrium
effects treated stochastically through additive Gaussian and
multiplicative power-law heat flow.  Our summary and conclusions are
contained in Sec.~\ref{sec_concl}.

\section{THERMALLY EQUILIBRATED SYSTEM AT REST }             
\label{sec_rest}

\subsection{Fluctuations}
We begin by considering a system of  
infinite spatial 3-volume $V_3$ at rest and in thermal equilibrium. 
The isospin 
charge operator, ${\calQ}^a$, is obtained from the zeroth component of the 
pertinent current, ${\calV}_\mu^a$, through
\be
{\calQ}^a (0)=\int\,\,d^3x\,{\calV}_0^a (0,\vec{x}) =\int\,\,d^3x\,\, 
\bar{q}(x)\,\gamma_0\,\frac{\tau^a}2\, {q}(x) \ ,  
\label{01}
\ee
where $q$ denotes quark-field operators and  
$\vec \tau$ are the standard $SU(2)$ isospin matrices 
(here and in what follows calligraphic letters indicate operators). 
The thermal fluctuations are defined by 
\ba
\Delta Q^2 &\equiv&
\left\langle {\calQ}^a (0) \,{\calQ}^a (0) \right\rangle_c
\nonumber\\
&=& \int d^3x d^3x' \left
\langle {\calV}_0^a (0,x)\,{\calV}_0^a (0,x') \right\rangle_c
\nonumber\\
&=& V_3 \int d^3x \left\langle {\calV}_0^a (0,x)\,{\calV}_0^a(0,0) 
\right\rangle_c
\nonumber\\
 &=&V_3 \ \frac{1}{\beta} \ i \int d^3x \int dx^0 \theta( x^0) 
\left\langle [{\calV}_0^a (x), {\calV}_0^a(0) ]\right\rangle \,,   
\label{QQcor}
\ea
where 
\be
\langle \calO \rangle \equiv 
\Tr \left[\e^{-\beta (\calH-\mu\calQ-\Omega)} \calO \right]  
\ee
indicates the thermal expectation value of the operator $\calO$.  The
subscript $c$ in $\langle \cdots \rangle$ refers to connected
parts. The relevant chemical potential is denoted by $\mu$,
$\beta=1/T$ is the inverse temperature, and $\Omega$ is the
thermodynamic potential.  The last line in Eq.~(\ref{QQcor}) follows
from general analytic properties of the retarded correlator with the
order of integration fixed to enforce approach from the space-like
regime.

\subsection{Susceptibilities}

The local fluctuations in a globally conserved quantity can be
expressed through a derivative of the thermal expectation value of $\calO$ with
respect to the associated chemical potential (see, \eg,
Refs.~\cite{LLSM,DS}):
\be
(\Delta \calO)^2 = \la \calO^2 \ra -\la \calO \ra^2 =
 - T \frac{ \partial \la {\cal O} \ra}{\partial \mu} \ .
\ee
Since the charge-density is defined as a partial derivative 
of the thermodynamic potential, its thermal fluctuations are 
given by the charge susceptibility 
\be
\chi_{ch} \equiv \frac{1}{V_3} \frac{\partial^2 \Omega}{\partial \mu_{ch}^2}  
= \frac{(\Delta Q)^2}{T V_3} \ .
\ee

A general relation between the susceptibility and the $00$-component
of the electromagnetic (e.m.) polarization tensor is (see, \eg,
Refs.~\cite{McL87,Kapu,PZ,BIP95}) 
\be
\chi_{ch}  \equiv -\Pi^{00}(q_0=0,\vec q\to 0) =  m_D^2  \ , 
\label{m_D}
\ee 
where $m_D$ is the Debye mass (strictly speaking, this relation is valid 
only to leading order in $e^2$). Equation~(\ref{m_D}), also implied by
Eq.~(\ref{QQcor}), then provides a direct link to the electric 
screening length in the system. On kinematical grounds, one expects
$m_D^2(T)$ to receive contributions chiefly from low-lying mesonic
excitations in a hadronic gas and from (nearly) massless quarks and
anti-quarks in a partonic gas.

\subsection{Examples in Hadronic and Quark-Gluon-Plasma Phases}
\label{sec_2C}
We turn now to estimates of charge fluctuations assuming thermal
equilibrium (this assumption is relaxed later in Sec.~\ref{sec_offeq}).  
A meaningful quantity accessible to experiments is
the magnitude of the hadronic fluctuations normalized to the entropy
$S$ in a suitable subvolume $V_3$ of the system (realized, \eg, via a
restricted rapidity interval of data):
\be
R=\frac{(\Delta Q)^2}{S}=\frac{(\Delta Q)^2/V_3}{S/V_3}= \frac{Tm_D^2}{s} \ , 
\ee  
where $s$ is the entropy-density. In the ideal gas approximation,
\be
(m_D^0)^2 = \beta \sum\limits_i d_i^{ch} \int 
        \frac{d^3k}{(2\pi)^3} \ f(\omega_i) \ [1\mp f(\omega_i)]
\label{mD0} 
\ee
and
\ba
s = \mp  \sum\limits_i d_i \int   \frac{d^3k}{(2\pi)^3}
[&\pm& f(\omega_i) \ln f(\omega_i) 
\nonumber \\
&+& (1\mp f(\omega_i)) \ln (1\mp f(\omega_i)] \,,
\label{s0}
\ea
where the summations run over all particle species $i$; the upper
(lower) sign refers to fermions (bosons), $d_i^{ch} \equiv d_i^{spin}
\ \sum_{I_3} (q_i^{I_3})^2$ denotes the (squared) charge degeneracy 
of a given isospin-multiplet, $d_i$ is the corresponding total
degeneracy, and $\omega_i^2=m_i^2+k^2$. The thermal distribution
functions $f=f^B$ or $f^F$ correspond to bosons or fermions,
respectively.  It is important to note that, despite their simplicity,
Eqs.~(\ref{mD0}) and (\ref{s0}) take account of interactions between
the hadrons to the extent that the thermodynamics of an interacting
system of hadrons can be approximated by that of a mixture of ideal
gases of ``elementary'' particles and their
resonances~(cf. \cite{virial} and references therein).

For a massless pion gas, $R_\pi^0 = 0.253$.  With the empirical vacuum
pion mass, $R^{id}_\pi (T_{chem}) \cong 0.18$ at a typical chemical
freeze-out temperature $T_{chem} \cong 180 $ MeV.  As pointed out in
Ref.~\cite{AHM00}, inclusion of higher-mass hadronic resonances
reduces this number further to $R^{id}_{HG} \cong 0.06-0.07$.  This is
readily understood from the fact that heavier particles give
relatively larger contributions to the entropy density than they do to
the susceptibility (see Eqs.~(\ref{mD0}) and (\ref{s0})). 

In contrast, in a non-interacting QGP with 2 massless flavors and $N_c$ 
colors, $R_{QGP}$=$\frac{1}{3} N_c \frac{5}{9}/(37\times 4\pi^2/90)$=0.034.
Perturbative calculations to higher order in the strong coupling constant 
can be found, \eg, in Ref.~\cite{LB96}.  The use of
nonperturbative results from lattice gauge simulations~\cite{Go97} 
to extract information on the electromagnetic
susceptibility has been alluded to, \eg, in Ref.~\cite{JK00}.  As
expected, the fluctuations in a QGP at temperatures $T\gg T_c$ seem to
be rather close to the naive perturbative result. Close to $T_c$, the
situation is less clear.

\subsection{Average Phase Space Densities} 
Liouville's theorem guarantees that the final state average phase space
density of a system gets frozen at a certain time and gives a measure
of the dynamics in the prior interacting stage.  As observed in 
Ref.~\cite{Ber94}, interferometric (Hanbury-Brown and Twiss)
studies of particle correlations provide a measure of the average
phase space density of the particles in the final state.  We add here
that such knowledge also provides for a determination of the Debye mass.
This is easily seen by recasting Eq.~(\ref{mD0}) for the
squared Debye mass to read as
 \ba
(m_D^0)^2 &=& \frac {n_{ch}}{T} \left[ 1 \mp \langle f
\rangle_{ch} \right] \,, \quad {\rm or} 
\label{phspace1} \\ 
\langle f \rangle_{ch} &=& 
\mp \left [ \frac {T(m_D^0)^2}{n_{ch}} - 1 \right] \,, 
\label{phspace2}
\ea 
where $n_{ch}$ is the total number density of charged particles and
$\langle f \rangle_{ch}$ is the average phase space density of all
charged particles.  To the extent that $n_{ch}, T$, and $\langle f
\rangle_{ch}$ can be determined from experiments, at least at the
dilute freeze-out stage of a heavy-ion collision, an estimate of the
Debye mass is rendered possible from Eq.~(\ref{phspace1}).
Equation~(\ref{phspace2}) is also useful, insofar as it provides a
consistency check on models that predict the Debye mass from first
principles~\cite{EVK93}. 

For a single species of bosons or fermions, 
the average phase space density may be expressed as
\ba
\langle f \rangle = \int   \frac{d^3k}{(2\pi)^3} \, f^2 \bigg/
 \int   \frac{d^3k}{(2\pi)^3} \,f  \qquad \qquad 
\nonumber \\
= \frac { \sum\limits_{j=1}^\infty (\pm )^{j+1}\left(\frac{j}{j+1}\right) 
\exp[(j+1)\beta\mu]~ K_2[(j+1)\beta m] }
{ \sum\limits_{j=1}^\infty(\pm )^{j+1}\left(\frac{1}{j}\right) 
\exp[j\beta\mu]~ K_2[j\beta m] } ,
\label{eph}
\ea
where the $+(-)$ sign refers to bosons (fermions) and 
$K_2(x)$ is a modified Bessel function. For temperatures such
that $T/m \ll 1$, use of the asymptotic relation $K_\nu(x) \sim
(\pi/2x)^{1/2}~\exp(-x)$ reduces Eq.~(\ref{eph}) to the result 
\ba \langle
f^{B,F}\rangle \simeq \frac{\exp[\beta(\mu-m)]}{(2\sqrt 2)}\,, 
\ea 
which is valid in the classical regime.

For orientation, we show in Fig.~\ref{phs_fig} results for the
average phase space density $\langle f^B\rangle_{ch}$ of charged pions
and kaons for selected values of their respective chemical potentials
as functions of temperature.  
Although results for temperatures up to 200 MeV are shown in this figure 
for illustrative purposes, we should focus our attention on  
temperatures below $T_c$, the temperature at which the
transition to a QGP phase occurs. For reference, the dotted line in the
upper panel shows the result $\langle f^B \rangle_{ch}\simeq 0.37$ for
massless pions at zero chemical potential (in contrast, $\langle f^F
\rangle_{ch}\simeq 0.088$ for massless fermions at zero chemical
potential).  All other curves show results with vacuum values of
masses. The dashed curves in both panels are for zero chemical
potentials.  Evident features from the results of this figure are: (1)
Increasing masses of bosons reduce $\langle f \rangle_{ch}$; (2)
Positive (negative) chemical potentials of bosons enhance (suppress)
$\langle f^B\rangle_{ch}$ relative to its value at $\mu=0$. For
chemical potentials approaching the mass of the particle (incipient
Bose condensation), the average phase space occupancy begins to become
increasingly large; and (3) The higher (lower) the temperature, the
higher (lower) is $\langle f^B\rangle_{ch}$. 
\bfg[!t]
\bce
\epsfig{file=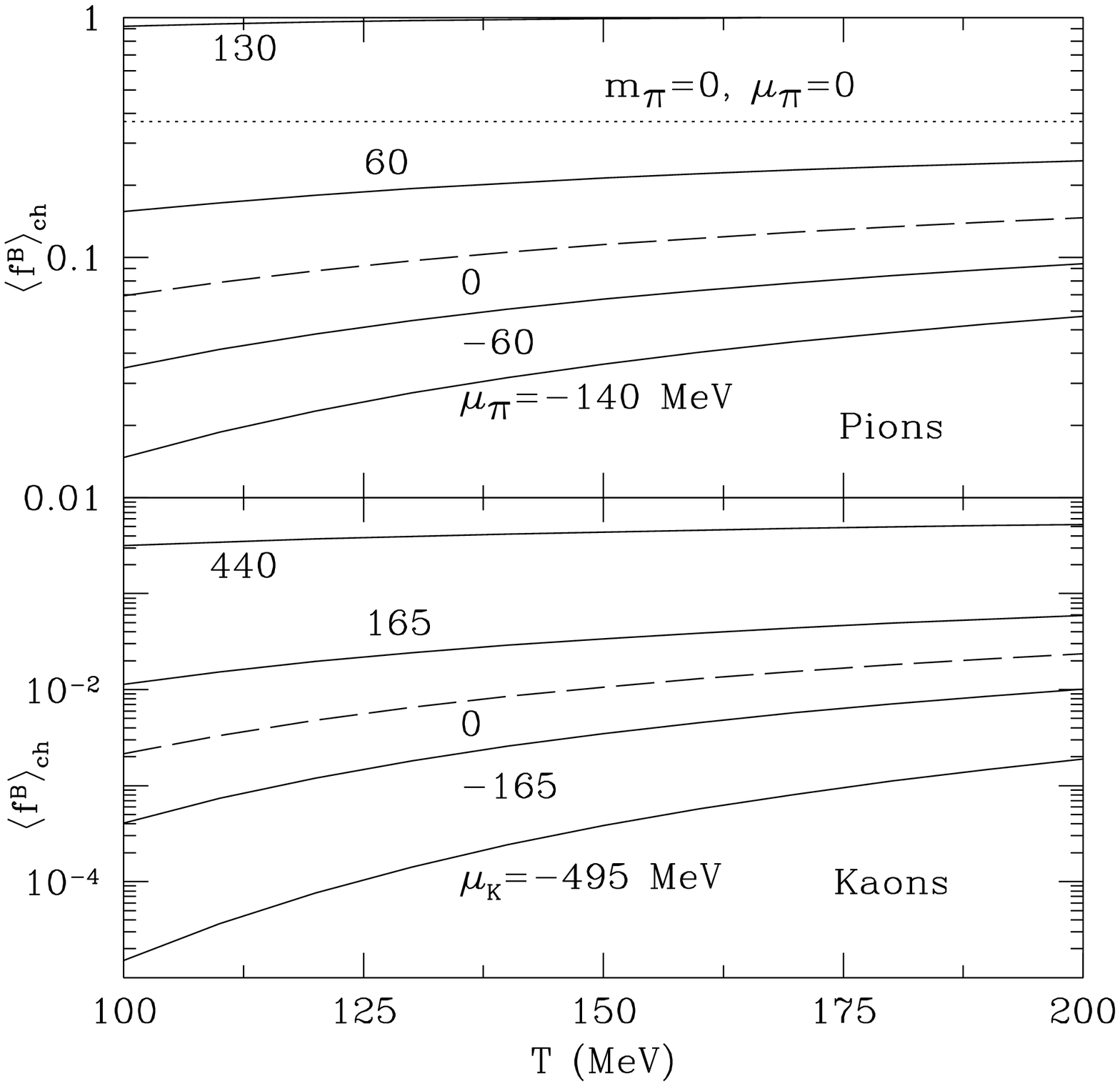,width=8cm,angle=0}
\ece
\caption{Average phase space densities $\langle f^B\rangle_{ch}$ of
charged pions (upper panel) and kaons (lower panel) as functions of
temperature at the indicated values of their respective chemical
potentials. The dotted curve in the upper panel shows $\langle
f^B\rangle_{ch} \simeq 0.37$ for massless pions at zero chemical potential.}
\label{phs_fig}
\efg

It is straightforward to compute $\langle f\rangle_{ch}$ in a mixture
of hadrons through the use of Eq.~(\ref{eph}) by summing over the
various species with their appropriate charge
degeneracies $d_i^{ch}$.  We refer the reader to
Refs.~\cite{Ber94,Fer99}, where the extent to which the conditions at
the freeze-out stage of the collision can be delimited are discussed.

\subsection{Electromagnetic Correlations}

An intimate connection exists between the fluctuations of the charged
constituents in the system and the emission of electromagnetic
radiation.  The e.m. polarization tensor figuring in Eq.~(\ref{m_D})
directly governs the production rates of dileptons and photons.  This
is important, since electromagnetic emission occurs throughout the
evolution of the system created in heavy-ion collisions and ranks
highly among the principal ways in which the dense stages of evolution
can be directly probed (cf.~\cite{RW00} and references therein).

Beginning with dilepton ($l^+l^-$) emission, we note that  
the rate of production from a thermally equilibrated system is given by 
\ba
\frac{dR_{l^+l^-}}{d^4q}&=& L^{\mu\nu}(q) \ W_{\mu\nu}^{\em}(q)
\nonumber\\ &=& -\frac{\alpha^2}{\pi^3 M^2} f^B(q_0;T) \frac{1}{3}
\left({\rm Im}\, \Pi^L + 2 {\rm Im}\, \Pi^T\right) \,,
\label{rates}     
\ea 
where $L^{\mu\nu}$ is the usual lepton tensor, and the electromagnetic 
hadron tensor 
\be
W_{\mu\nu}^{\em}(q)=
\int\,d^4x\,\e^{iq\cdot x}\,\left\langle
{\cal J}_\mu^{\em} (x) \,\,{\cal J}_\nu^{\em} (0) \right\rangle_c
\ee
is given directly in terms of the imaginary part of the retarded 
photon self-energy
\be
W_{\mu\nu}^{\em}=-2~f^B(q_0;T)~{\rm Im}~\Pi_{\mu\nu} \, . 
\ee

Photon emission rates are obtained by taking the limit 
$M^2\equiv q_0^2-q^2 \rightarrow 0$ in the above expressions.  Note that, 
unlike the space-like limit involved in Eq.~(\ref{m_D}), dilepton (photon)
production probes the time-like regime (limit), as well as both the
longitudinal and transverse parts of the electromagnetic polarization
tensor.

We turn now to elaborate on this interrelation, beginning with the simple
and transparent case of an ideal gas of massless constituents (either
pions or quarks). In this case, the leading temperature contribution
to the one-loop self-energy is (see, \eg, Ref.~\cite{LB96})
\ba
\Pi^{00}(q_0,q) &=& -m_D^2 \  d_Q  
\nonumber\\
 &\times& \left[ 1-\frac{q_0}{2q}
\left(\ln\left|\frac{q_0+q}{q_0-q}
\right| - i \pi\theta(-M^2)\right)  \right]   \ , 
\label{PI_space}
\ea
where $d_Q=1$  ($N_c \sum_q e_q^2$) for pions (quarks).  
Note that $\Pi^{00}(q_0,q)$   
acquires an imaginary part only for space-like $M^2<0$, 
due to Landau damping.
The dilepton production rate, which stems from 
the time-like region characterizing decay contributions (emissivities), 
takes the form 
\ba
\frac{dR_{l^+l^-}}{d^4q} &=& \frac{\alpha^2 \tilde d_Q}{\pi^3 M^2}
 f^B(q_0) \  \frac{M^2}{48\pi} \left[1+\frac{2T}{q}  
\ln\left(\frac{x_+\mp 1}{x_-\mp 1}\right) \right]  
\nonumber\\  
 &\stackrel{q\to 0}{=}& \frac{\alpha^2 \tilde d_Q}{\pi^3 M^2} f^B(q_0) \ 
 \frac{M^2}{48\pi} \left[1 \pm 2f\left(\frac{q_0}{2}\right)\right] \ ,   
\ea 
where $\tilde d_Q=1$ for pions or $N_s^2 N_c \sum_q e_q^2$ for quarks
($N_s=2$), and $x_\pm=\exp[-(q_0\pm q)/2T]$.

For illustrative purposes, the $q\to 0$ limit has been taken in the rightmost
expression.  Note that the overall Bose factor is not part of the
photon self-energy so that the leading temperature contribution to
dilepton production is due to the imaginary part of the vacuum
piece. In the $q\to 0$ limit, the explicit form of the 00-component for 
time-like $M^2>0$ is
\be
{\rm Im}\, \Pi^{00} (q_0,q)=- \frac{q^2}{12\pi} 
\left[ 1 \pm  2f\left(\frac{q_0}{2}\right) \right]\ .  
\label{PI_time}
\ee 
A formal connection between the space-like and time-like regimes can, in
principle, be constructed via a dispersion relation. For the case of
interest here, the analytic properties of the retarded photon
self-energy imply that
\ba 
{\rm Re}\, \Pi^{00}(q_0=0,q)= -\int\limits_0^\infty
\frac{d\omega^2}{\pi} \frac{{\rm Im}\, \Pi^{00}(\omega,q)}{-(\omega^2)}
\qquad\qquad \nonumber\\ \quad 
= -\int\limits_0^q
\frac{d\omega^2}{\pi} \frac{{\rm Im}\, \Pi^{00}(\omega,q)}{-(\omega^2)}
- \int\limits_q^\infty \frac{d\omega^2}{\pi} \frac{{\rm Im}\,
\Pi^{00}(\omega,q)}{-(\omega^2)} \ ,
\label{disp}
\ea 
where the second equality explicitly shows the decomposition into
space-like and time-like contributions. Using the imaginary parts from
Eqs.~(\ref{PI_space}) and (\ref{PI_time}), we find that ${\rm Re}\,
\Pi^{00}(q_0=0,q)$, and hence the Debye mass, is saturated by the
space-like part of the dispersion integral, the second term in
Eq.~(\ref{disp}) giving a vanishing contribution.  Thus, for a
thermally equilibrated system at rest, a model-independent relation
between fluctuations and dilepton production cannot be achieved.
However, within a given approach for calculating electromagnetic
emissivities, the Debye mass and hence charge fluctuations are 
simultaneously determined.

\subsection{Effects of Interactions on Electromagnetic Correlations}
\label{sec_int}
The interrelations between fluctuations, susceptibilities, and the
photon polarization function described in the previous sections set
the stage to address the effects of interactions between the
constituents in a thermal equilibrium system.  In a hadronic gas at a
finite temperature and baryon chemical potential, various approaches
to calculate $W^{\mu\nu}$ have been
pursued~\cite{RW00,GK91,RCW97,SYZ97,RW99,SZ}.  In the following, we employ two
examples related to dilepton and photon emissivities  
to analyze the impact of interactions on fluctuations.  

\subsubsection{Vector Dominance Model}

A number of hadronic models used to study medium modifications invoke
the phenomenologically successful Vector Dominance Model (VDM) at
finite temperatures and densities using effective hadronic
Lagrangians~\cite{GK91,RCW97}.  In VDM, the photon couples to charged
hadrons exclusively through the vector mesons $\rho$, $\omega$ and
$\phi$. In the isovector $\rho$-channel, which constitutes the dominant
part of the photon self-energy, the latter then
takes the form
\be
\Pi^{\mu\nu}(q_0,q;\mu_B,T)= \frac{\mrhozf}{g_\rho^2}
D_\rho^{\mu\nu}(q_0,q;\mu_B,T) \,.  
\ee 
With the help of standard projection operators $P_L^{\mu\nu}$ and 
$P_T^{\mu\nu}$~\cite{LB96},  the in-medium $\rho$ propagator can be 
decomposed into its longitudinal ($L$) and transverse ($T$) parts:  
\ba
D_\rho^{\mu\nu}(q_0,q;\mu_B,T )&=& D_\rho^L P_L^{\mu\nu} + D_\rho^T
P_T^{\mu\nu} \nonumber\\
D_\rho^{L,T}&=&\frac{1}{M^2-\mrhozt-\Sigma_\rho^{L,T}} \ . 
\ea 
Above, $g_\rho$ denotes the universal VDM coupling constant, $m_\rho^{(0)}$
is the bare $\rho$ mass, and $\Sigma_\rho^{L,T}$ stands for all possible
proper self-energy insertions. 

Let us first focus on the $\rho\to\pi\pi$ loop contribution to
$\Sigma_\rho$, which involves the strong coupling of the $\rho$.  The 
resummation of this loop to all orders via the propagator accounts for
interactions, albeit from a select class of diagrams.  To extract the
electromagnetic Debye mass, we calculate the $\rho\pi\pi$ loop at finite 
temperature~\cite{GK91}, take the space-like limit and, after
subtracting the vacuum piece, obtain
\ba
m_D^2 &=&-\frac{\mrhozf}{g_\rho^2}  \lim\limits_{q\to 0} 
\left[D_\rho^{00}(0,q) - D_{\rho,vac}^{00}(0,0) \right]
\nonumber\\
     &=& (m_D^0)^2 \frac{\mrhozt}{\mrhozt + \tilde m_D^2} 
\label{mDvdm}
\ea    
with $(m_D^0)^2$ given by Eq.~(\ref{mD0}).  The quantity $\tilde
m_D^2=g_\rho^2~(m_D^0)^2$ represents the ``strong'' (isovector) Debye
mass induced by the $\rho\pi\pi$ loop.  The rescattering of pions
through the formation of the $\rho$ resonance causes a reduction of
the Debye mass; in turn, the hadronic charge fluctuations are reduced.
This result is shown by the solid line in Fig.~\ref{fig_vdmcorr},
where at temperatures close to $T_{chem}$ a reduction of $\sim 25\%$
is apparent. If one accounts for effective pion-number conservation
(which is implied by chemical freezeout in a heavy-ion collision), the
resummation effect is more pronounced at lower temperatures.  At SPS
energies, pion-number conservation entails the build-up of a finite
pion chemical potential which can reach values $\mu_\pi\simeq 70$~MeV
for thermal freezeout at $T_{fo}\simeq 120$~MeV~\cite{RW99}.  This
effect is illustrated by the dashed curve in Fig.~\ref{fig_vdmcorr},
which shows that even around thermal freezeout
a reduction of $\sim 20\%$ of the fluctuation content persists.

Contributions from heavier states in a hot mesonic environment are
also present, but are less significant. The $K\bar K$ loop enters
predominantly into the $\phi$ meson component of VDM (OZI rule), and
the $\rho\rho$ loop is kinematically suppressed. For typical
$T_{chem}$, we find
$(\tilde{m}_{D,\rho\rho}/\tilde{m}_{D,\pi\pi})^2$$\simeq$1/4, which
when resummed results in additional reduction, but only at at the few
percent level.  Potentially more important are ``off-diagonal''
interaction terms, most notably thermal $\pi a_1(1260)$ loops which
are induced by resonant $\rho$ scattering off thermal pions, or, if
abundant, baryonic excitations.  Various interaction vertices have
been employed in the literature, see, \eg, Ref.~\cite{GG98} for a
recent survey. It turns out that the frequently employed $\rho$-meson
tensor ($\rho^{\mu\nu}$) couplings, which have the advantage of 
individually maintaining gauge invariance, lead to a vanishing 
contribution to the electromagnetic Debye mass. However, other 
choices, which do not vanish in the space-like screening limit 
are possible.
\bfg[!ht]
\vspace{-0.8cm}
\bce
\epsfig{file=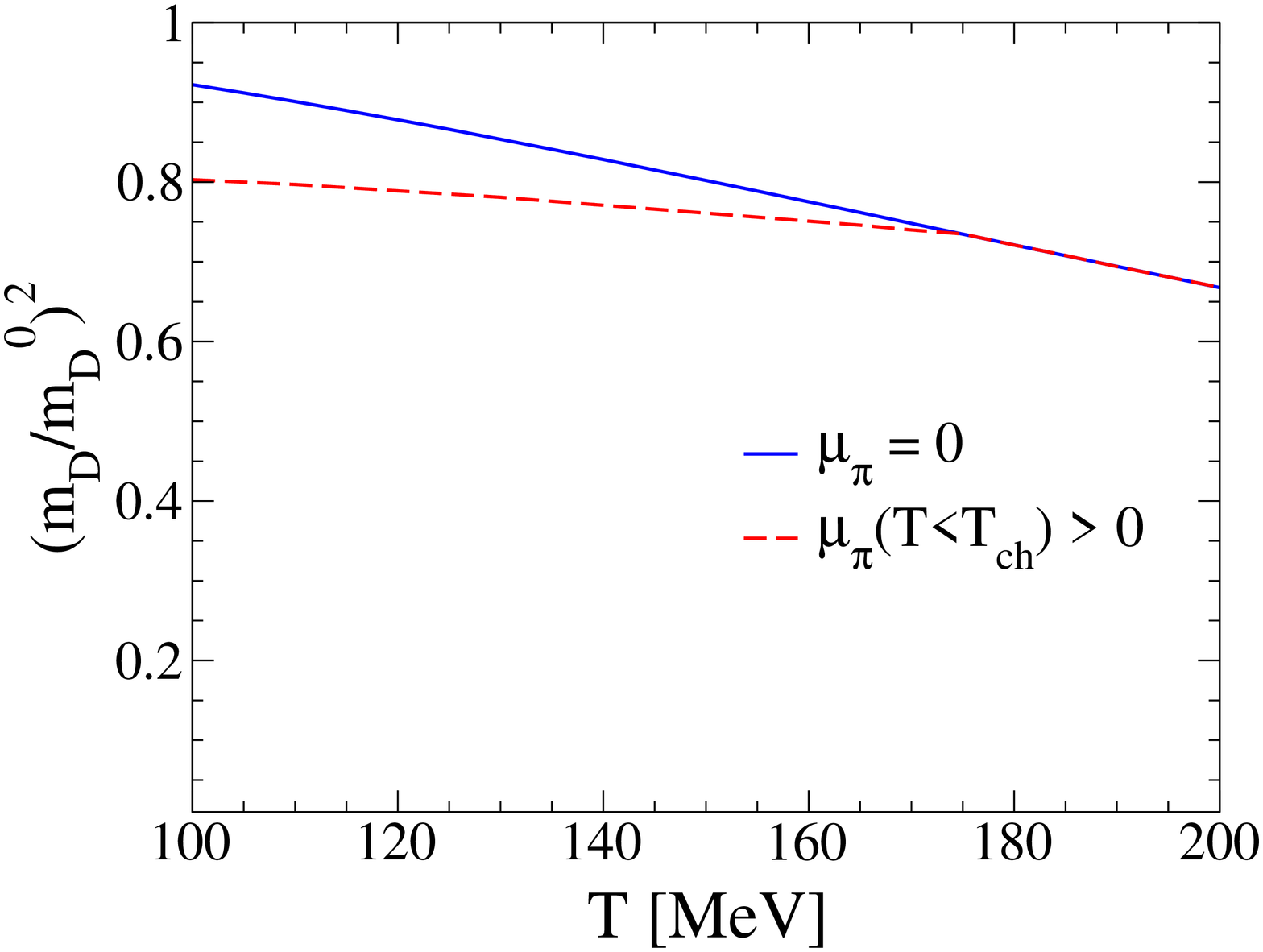,width=6.6cm,angle=-90}
\ece
\caption{VDM resummation correction to $\Delta Q^2$ represented by  
the ratio of the squared electromagnetic Debye mass that includes
effects of resummed $\rho\pi\pi$ loops (according to
Eq.~(\protect\ref{mDvdm})) to that from the lowest-order result in
strong interactions.}
\label{fig_vdmcorr}
\efg

\subsubsection{Chiral Reduction Scheme} 
The framework of Ref.~\cite{SYZ97} allows chiral master formulae to be used 
to relate $W_{\mu\nu}$ to experimentally extractable spectral functions 
in a model independent way. 
Using the analytical properties of ${W}_{00}$ in the complex
$\omega$-plane, it follows that, in thermal equilibrium,
\ba
{W}_{00}^{ab} (\omega,q) 
&=& - \frac {2\,{\rm Im}\,\Pi_{00}^{F, ab} (\omega,q)}
{\e^{\beta |\omega|} +1}
\nonumber\\
&=& - \frac {2\,{\rm Im}\,\Pi_{00}^{R, ab} (\omega,q)}
{\e^{\beta \omega} -1}  \ ,  
\label{Wab2}
\ea
where the superscript $F$ ($R$) explicitly indicates the Feynman 
(retarded) charge-charge correlator.  In this scheme, the vacuum part 
${\Pi}^{0F}$ is common to a hadronic or partonic gas and is related to 
experimentally measured spectral functions. One has  
\be
{\Pi}_{00}^{0F,ab} (M^2) =
i\int\,d^4x\,e^{iq\cdot x}\,\langle 0|
T^* \,{\calV}_0^a (x) \,\,{\calV}_0^b (0) |0\rangle_c \,,
\label{W0Fab}
\ee
which is given  by the longitudinal part of the isovector
correlation function,  
\be
i\,\delta^{ab}\,(g_{00}\,M^2-q_0^2 )\, \Pi_V^L(M^2)\,\,.
\label{PiV}
\ee
In a hadronic gas, the corrections to the leading spectral density
follow from a virial expansion. Specifically,
\ba
&&{\Pi}_{00}^{1F,ab} (\omega,q) =
\sum_A\int\frac{d^3k}{(2\pi)^3}\, \frac{f^\pi(\omega_k)}{2\omega_k}\,
\nonumber\\
&&\times
i\int\,d^4x\,e^{iq\cdot x}\,\langle \pi (k)|
T^* \,{\calV}_0^a (x) \,\,{\calV}_0^b (0) |\pi (k)\rangle_c\nonumber\\
&& + ... \,,
\label{W1Fab}
\ea
where $\omega_k^2=m_\pi^2+{\vec k}^2$.
The ellipsis refer to corrections of higher order 
in pion- and nucleon-densities.
The reduction of the forward Compton amplitude in Eq.~(\ref{W1Fab})
involves both the vector and axial-vector correlation functions.
One finds
\begin{eqnarray}
&&{\rm Im}\,{\Pi}_{00}^{1F, ab} (\omega,q) \approx\delta^{ab}\,
\int\frac{d^3k}{(2\pi)^3}\, \frac{f^\pi(\omega_k)}{2f_\pi^2\,\omega_k}\,
\nonumber\\
&&\times \bigg[-4\vec{q}^{~2} \,{\rm Im}\,{\Pi}_V (q) \nonumber\\&&\qquad+
2(\vec{k}+\vec{q})^2\,{\rm Im}\,{\Pi}_A (k+q) + (q\rightarrow -q)
\nonumber\\&&\qquad
+4(2\omega_k+\omega)\, (\omega_k\vec{q}^{~2} -\omega\,\vec{k}\cdot\vec{q})\,
\nonumber\\   
&&\qquad\times
{\rm Im}\,{\Pi}_V (q) \,{\rm Re}\,(\Delta_R(k+q) + q\rightarrow -q)
\bigg] \,,
\label{14x}
\end{eqnarray}
where $f_\pi$ is the pion decay constant and  $\Delta_R(k)$ is the 
retarded pion propagator. In obtaining this result,   
only the dominant contributions to the Compton amplitude
were retained. We note that the current algebra result follows
from setting $k=0$, as a result of which the quantity in brackets in 
Eq.~(\ref{14x}) takes the form
\be
4\vec{q}^{~2}\, \left( {\rm Im}\,{\Pi}_A (q) -
\,{\rm Im}\,{\Pi}_V (q) \right)\,\,.
\label{14y}
\ee
This relation illustrates the role of chiral symmetry restoration at
next-to-leading order in the hadronic (pionic) density.  Higher order
corrections, not discussed here, are calculable in a similar way and
in close analogy with calculations of photon and dilepton emissivities.

\section{CONNECTIONS TO HEAVY-ION EXPERIMENTS } 

Since fluctuations at the dilute freeze-out stage of a heavy-ion
collision have been amply discussed in the literature, we focus here
on how one might gain knowledge about fluctuations during the dense
stages of the collision.  Direct photons and dileptons, due to their
negligible final state interaction in a heavy-ion collision, have long 
been recognized as sensitive probes of the physical state of strongly
interacting matter.  We are thus led to the following natural question: 
To the extent that electromagnetic observables at the CERN-SPS have 
been adequately described through a combination of hadronic and partonic 
approaches~\cite{RW99,SZ,PP} (see \cite{RW00} for a review), what are the 
implications of dilepton and photon spectra to be measured at RHIC and LHC 
on charge fluctuations of hadrons and partons during 
the dense stages of evolution?  The answer to this question hinges
critically on the outcome of upcoming measurements and on the ability
of theory to model the complex evolution of the system created in a
heavy-ion collision.  In the following section, we provide a suitable
theoretical framework that can be systematically improved.
 
\subsection{Longitudinally Expanding System in Local Thermal Equilibrium} 
\label{sec_moving}
In a heavy-ion collision, isospin fluctuations occur
in an expanding environment that initially consists of partons which
hadronize towards the end of evolution. In the Bjorken expansion
scenario with boost invariance~\cite{Bjo83}, the isospin charge per unit
pseudo-rapidity can be schematically defined as $dQ^a/d\eta$. Boost
invariance and charge conservation imply that 
\be 
\frac
{d(dQ^a/d\eta)} {d\tau}=0 \,.  
\ee 
This conservation law assures that the mean fraction of isospin is the
same in all frames, and constant throughout the proper time evolution.
If this were to hold for the variance, then we would naively conclude
that the rest frame results developed in the preceding section will
trivially extend to the rapidity-binned fluctuations in a heavy-ion
collision. Closer inspection reveals that this is not the case, as 
will be shown below. 

Consider the thermal fluctuations in an expanding fluid-element that
is characterized by its pseudo-rapidity $\eta=\tanh^{-1}\,(t/z)$ and
proper time $\tau=\sqrt{t^2-z^2}$. 
Following the standard generalization of a conserved current 
to a moving frame, we define the isospin charge operator  
\be
\calQ_\Delta^a (\tau, \eta) = \int d\sigma^\mu \ \calV_\mu^a  
\ee
in a fixed pseudo-rapidity interval of extent $\Delta$
centered around $\eta$. Above, $\sigma^\mu$ represents a 3-dimensional
hyper-surface in space-time which is orthogonal to the local fluid 4-velocity
$u^\mu=\gamma(1,\vec\beta)$.  
From here on,  we restrict ourselves to purely longitudinal 
expansion for the sake of simplicity. 
With $d\sigma^\mu= dx_\perp (dz,0_\perp,dt)$, the charge becomes
\be
\calQ_\Delta^a (\tau,\eta) = \int dx_\perp 
\int\limits_{\eta_-}^{\eta_+} \tau  d\ueta \ u_\|^\mu \, 
\calV^a_\mu\left(\tau\ch \ueta, x_\perp, \tau\sh\ueta\right)  \,,
\ee
where $\eta_\pm=\eta\pm\Delta/2$. The quantity 
$u_\|^\mu =(\ch\ueta,0_\perp,\sh\ueta)$ is equivalent to the 
longitudinal Lorentz boost operator $\Lambda_0^{~\mu}$, as it should.  
The binned isospin fluctuations can then be expressed as 
\ba
\langle \calQ_\Delta^a (\tau, \eta ) \ \calQ_\Delta^b (\tau,\eta) \rangle_c
=\int dx_\perp \ dx'_\perp
\int\limits_{\eta_-}^{\eta_+} \int\limits_{\eta_-}^{\eta_+}
\tau d\underline\eta \ \tau d\underline\eta'
\nonumber\\
\times \ u_{||}^\mu \ {u'}_{||}^\nu \
\langle\calV_\mu^a(x_{||},x_\perp)\,\calV_\nu^b (x_{||}',x_\perp')\rangle_c \,,
\label{QQeta}
\ea 
where $x_\|\equiv \tau(\ch\ueta,\sh\ueta)$. The 
averaging is carried in a hadronic or a partonic gas 
in equilibrium at temperature $T (\tau)=1/\beta (\tau)$ and pertinent chemical
potentials. (We will return to discuss some specific issues related to 
non-equilibrium situations  
in Sec.~\ref{sec_offeq}.) 
Using translational invariance in the transverse direction along with 
a Fourier transform of the unordered charge correlator, 
\be
{\bf W}_{00}^{ab} (\omega,q) = \int d^4x \ \e^{iq \cdot x} \
\langle \calV_0^a (x)~\calV_0^b (0) \rangle \ ,
\label{Wab}
\ee
Equation~(\ref{QQeta}) can be cast in the form 
\ba
&&\langle Q^a_\Delta (\tau, \eta ) \,\,Q_\Delta^b (\tau, \eta)\rangle_c = 
{V_\perp} \int\frac{d\bar\omega\,d\bar q_z}{(2\pi)^2}  \qquad\qquad
\nonumber\\
&&\times \ W_L^{ab}(\bar\omega/\tau,0_\perp,\bar q_z/\tau) \ 
\frac{|F_\Delta (\eta,;\bar\omega,\bar q_z)|^2}{\bar\omega^2-\bar q_z^2} \ .
\label{QQeta2}
\ea
Here, we have introduced dimensionless 
variables $\bar\omega =\tau \omega$ and $\bar q= \tau q$ and 
exploited the relations 
\ba
W_{00}=\frac{q^2}{M^2} W_L\,,~
W_{0z}=\frac{\omega q_z}{M^2} W_L\,,~{\rm and}~ 
W_{zz}=\frac{\omega^2}{M^2} W_L\,, \nonumber
\ea 
as required by current conservation.  
The kinematical ``formfactors'' in Eq.~(\ref{QQeta2}) arise from 
restricting the isospin charge to a bin of width $\Delta$ 
in pseudo-rapidity and are given by
\be
F_\Delta \ F_\Delta^* = \left[2~\sin\left(\sh\left(\frac{\Delta}{2}\right) \ 
[\bar\omega \, \sh\eta-\bar q_z \,  \ch\eta] \right) \right]^2 \ . 
\label{Fdelta}
\ee
The occurrence of solely the longitudinal part $W_L$ is a consequence
of the underlying assumption of purely longitudinal fluid flow and
current conservation.  The presence of transverse flow would naturally
involve the transverse part $W_T$ as well.

We recall that for dilepton emission, the combination $2~W_T+W_L$
contributes, whereas in photon emission only $W_T$ is
needed.  At large proper time $\tau$, the correlation function is
probed around $\omega\approx q\approx 0$.  In general, the fluctuations
in pseudorapidity bins represented by Eq.~(\ref{QQeta2}) are sensitive to
both space-like and time-like physics. As noted in the previous section,
the space-like contribution is driven by screening and Landau damping,
while time-like physics governs decays or production. The latter are
at the origin of photon and dilepton emissivities. To lowest order in
temperature, the imaginary part of the relevant polarization is
empirically accessible through $e^+e^-$ annihilation for $M^2>0$.
Since thermal fluctuations preserve isospin, the integrand in
Eq.~(\ref{QQeta2}) (modulo $F$) is a consequence of the fluctuation
dissipation theorem (cf.~Eq.~(\ref{Wab2})).

Employing boost invariance, we can focus on a window of width $\Delta$
centered around mid-rapidity $\eta=0$. Inserting 
Eq.~(\ref{Wab2}) into Eq.~(\ref{QQeta}), we obtain (omitting isospin
indices)
\ba
&&\frac{\langle \calQ_\Delta^2 (\tau) \rangle}{V_\perp} =
(-2) \, \left[2\, \sh(\Delta/2) \right]^2
\int\frac{d\bar\omega\,d\bar q}{(2\pi)^2} \qquad \quad
\nonumber\\
&& \qquad \times \  \frac{{\rm Im} W_R^{00}(\bar\omega/\tau,0_\perp,\bar q/\tau)}
{\e^{\beta\,\bar\omega/\tau}-1} \
\left(
\frac{\sin\left[\bar q\,\sh(\Delta/2)\right]} {\bar q\,{\rm sh}(\Delta/2)}
\right)^2  \ .
\label{Q2}
\ea 
This relation is one of the principal results of this work. It explicitly
demonstrates that fluctuations are sensitive not only to screening, but also to 
emissivities. Such a feature is in contrast to the case of the rest-frame
fluctuations discussed in Sec.~\ref{sec_rest},  and stems from
the general properties of boost transformations of charge-charge
correlators.

Equation~(\ref{Q2}) also shows that fluctuations are proportional to the
``size'' of the pseudorapidity interval through $\sh^2(\Delta/2)$,
vanishing for $\Delta=0$ as they should.  The dominant contribution
arises from long wavelengths $\lambda\simeq 2\pi/q \ge 2\pi\,\tau\,{\rm
sh} (\Delta/2)$ for which ${\rm sin}^2x/x^2\approx 1$, and small
frequencies $f=\bar\omega/2\pi\tau \simeq 1/2\pi\tau_{F}$, where
$\tau_{F}$ is a typical lifetime of the expanding system.  The
limitation on frequency is enforced by the requirement that for large
times, one has to recover the equilibrium screening limit that was
discussed in the previous section.  That this is indeed the case can
be seen as follows: For large $\tau$, the oscillating factor on the
right-hand-side of Eq.~(\ref{Q2}) collapses the $\bar q$ integration
to its saddle point at $\bar q \simeq 0$.  Furthermore, the
denominator becomes $\e^{\beta\,\bar\omega/\tau}-1 \simeq
\beta\bar\omega/\tau$. Using this, one recovers the dispersion
relation, Eq.~(\ref{disp}), for the Debye mass.  Note also that the
factor $z=2\tau\,\sh(\Delta/2)$ combines with $V_\perp$ to yield the
spatial 3-volume $V_3$.

Figure~\ref{fig_dQ2} displays results of the quantity $R= \langle
\calQ_\Delta^2 (\tau) \rangle/V_3 s(\tau)$ from numerical evaluations
of Eq.~(\ref{Q2}) and Eq.~(\ref{s0}) for a model system comprised of an
ideal QGP ($N_f=2$) in the partonic phase and massless pions in the
hadronic phase.
The time evolution of temperature has been taken to result from a
boost-invariant longitudinal expansion~\cite{Bjo83} in which
\ba
T(\tau) = 
    \left\{ \begin{array}{ll}  
 T_0~(\tau_0/\tau)^{1/3}  & , ~~~~\tau_0 < \tau < \tau_1  
\quad ({\rm QGP})   \\ 
 T_c  & , ~~~~  \tau_1 < \tau < \tau_2               \quad  ({\rm Mixed}) \\
 T_c~(\tau_2/\tau)^{1/3}  & , ~~~~ \tau_2 < \tau \qquad \quad \ ({\rm HG})
 \ .   \end{array}  \right. \,
\ea
The numerical values of the time constants, $\tau_0=1$~fm/c,
$\tau_1=2$~fm/c, $\tau_2=5$~fm/c and the critical temperature,
$T_c=170$~MeV, were chosen to generate conditions 
that resemble those at SPS energies 
(\eg, $T_0=T_c (\tau_1/\tau_0)^{1/3}\simeq 214$~MeV).  
In the mixed phase, the fluctuation content $R=f R_{QGP}
+(1-f) R_{\pi}$ is a weighted sum with the appropriate volume
fractions $f$ and $(1-f)$ of the QGP and hadronic phases,
respectively.
\bfg[!h]
\vspace{-0.7cm}
\bce
\epsfig{file=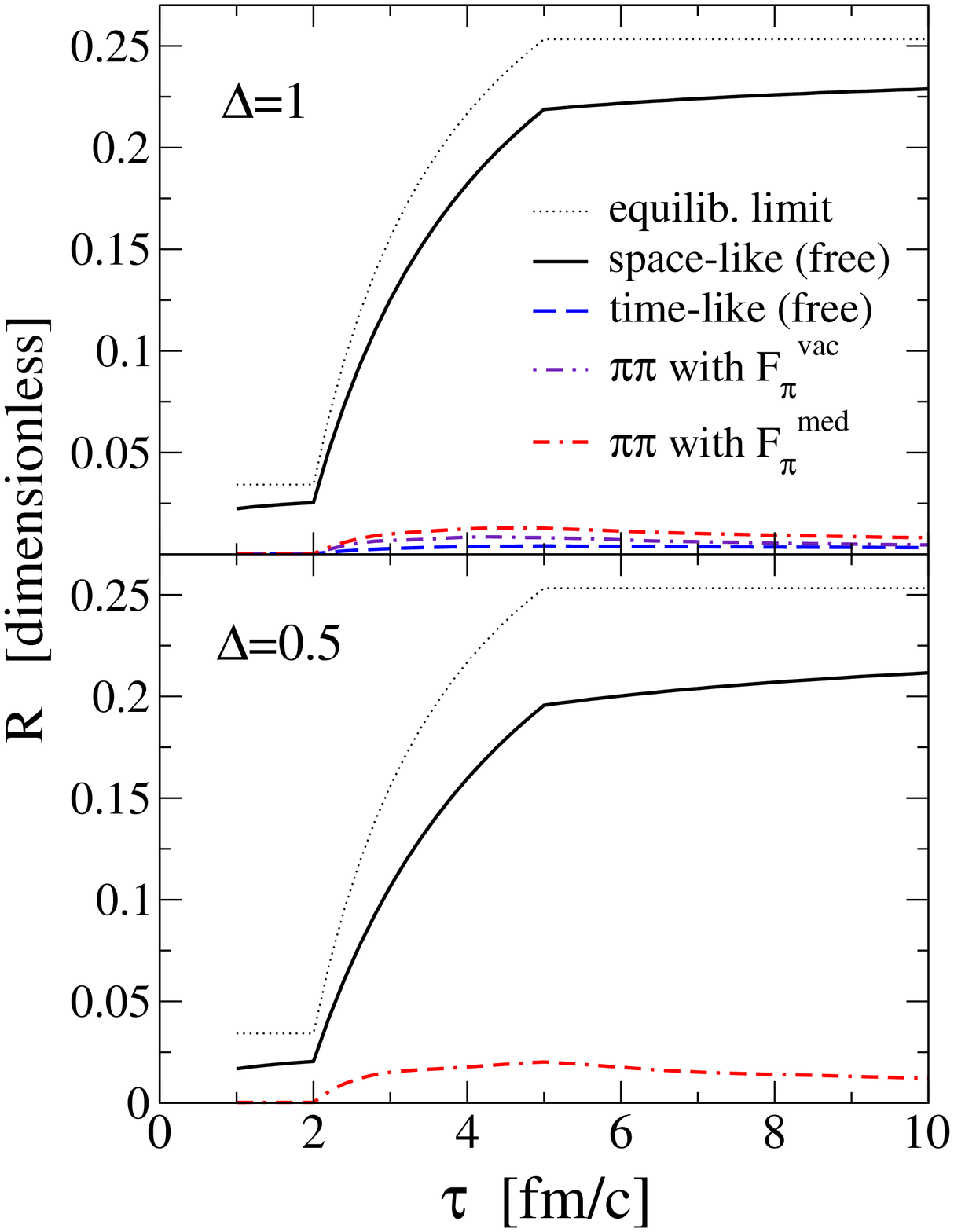,width=8.5cm}
\ece
\caption{Proper-time evolution of charge fluctuations per unit
3-volume, Eq.~(\protect\ref{Q2}), normalized to the appropriate
entropy density, Eq.~(\protect\ref{s0}), for an ideal QGP ($\tau \le
2$~fm/c) and a massless pion gas ($\tau\ge 5$~fm/c) in a locally
thermalized, longitudinally expanding fluid.  The upper (lower) panel
shows results for a pseudorapidity window of size $\Delta=1$ (0.5).
The solid (long-dashed) curves correspond to contributions from the
space-like (time-like) part of the charge-charge correlator.  The
dash-dotted and double-dash-dotted lines are results for the time-like
hadronic contribution wherein $\pi\pi$ correlations have been included
through the free pion-e.m.~formfactor $F_\pi^{vac}(M^2)$ and an
in-medium $F_\pi^{med}(M^2)$ (with twice larger $\rho$-width),
respectively.  The dotted lines indicate the static equilibrium limit,
$T m_D^2/s$.}
\label{fig_dQ2}
\efg

In the non-interacting case (employing for Im~$W^{00}_R$
Eqs.~(\ref{PI_space}) and (\ref{PI_time}) generalized to finite $q$),
fluctuations are clearly dominated by the space-like contribution
of the correlator (solid curves in Fig.~\ref{fig_dQ2}), especially at
small proper-times in the partonic phase. The latter fact is due to the 
fermionic character of the charge carriers (see Eq.~(\ref{PI_time})).  
However, as shown in Sec.~\ref{sec_int}, the inclusion of hadronic 
interactions tends to suppress the space-like contribution (not
shown in the figure), but enhances the time-like one, as is evident
from the dash-dotted and double-dash-dotted curves in the upper
panel.  This behavior becomes more pronounced if the size of the
rapidity bin is reduced (see the lower panel of Fig.~\ref{fig_dQ2}).

In general, $\Delta$ should be large enough to enhance the
volume-to-surface effect, but small enough to avoid the constraint of
overall charge conservation in the nuclear system.

We emphasize that the evolution of fluctuations shown in
Fig.~\ref{fig_dQ2} refers to a highly idealized model, especially in
the hadronic phase.  As stressed in Sec.~\ref{sec_2C}, the entropy
density of a more realistic hadronic resonance gas will be several
times larger than that of a massless pion gas~\cite{AHM00}, entailing
a reduction in $R$. To illustrate its effect, we show in
Fig.~\ref{fig_dQ2s} the results of a calculation in which the pure
hadronic phase begins with an entropy density $s \simeq 5~{\rm
fm}^{-3}$, a value representative of a hadronic resonance gas at
$T_c=170~\rm{MeV}$.  Finite hadron masses also imply that $s_{HG}$
falls off faster than $T^3$ with decreasing temperature, which has
been simulated by employing
$s_{HG}(\tau)=s_{HG}(T_c)~(\tau_2/\tau)^{1.35}$, indicated by thermal
fireball calculations~\cite{RW99}. Consequently, the static limit of
the ratio $R$ increases with falling temperature in the expanding
resonance gas for $\tau>5$~fm/c (see the dotted lines in
Fig.~\ref{fig_dQ2s}).  To minimally account for the effect of
additional states in the numerator of $R$ (this contains the magnitude
of fluctuations), we have also included contributions from 
kaons to the hadronic
e.m.~correlator (in the massless limit).  This amounts to a simple
increase of $\langle \calQ_\Delta^2 (\tau) \rangle/V_3$ by a factor of
2 compared with the result for massless pions. The presence of
massless strange quarks in the QGP hardly affects the equilibrium
value of $R_{QGP}$, reducing it from 0.034 (for $N_f=2$) to 0.032 (for
$N_f=3$). The relative role of space- and time-like contributions to
$R$ does not change as compared to Fig.~\ref{fig_dQ2}.

More realistic hadronic correlators, as were found necessary to explain the 
dilepton data from heavy-ion collisions at SPS energies, are characterized by a
substantial reshaping of the pion e.m.~form factor,  in particular
through an appreciable accumulation of strength at low invariant masses.  
We anticipate that the use of such correlators in
Eq.~(\ref{Q2}) will enhance the role of the time-like region.
Results for the simplified cases considered here are thus to be regarded
as suggestive.  They clearly indicate, however, that calculations of a
more realistic evolution of $R$ employing improvements in (i) the
equation of state, (ii) space-time evolution, and (iii) appropriate
electromagnetic correlators are worthwhile. Developments in this
regard are in progress and will be reported separately.
\bfg[!h]
\vspace{-0.8cm}
\bce
\epsfig{file=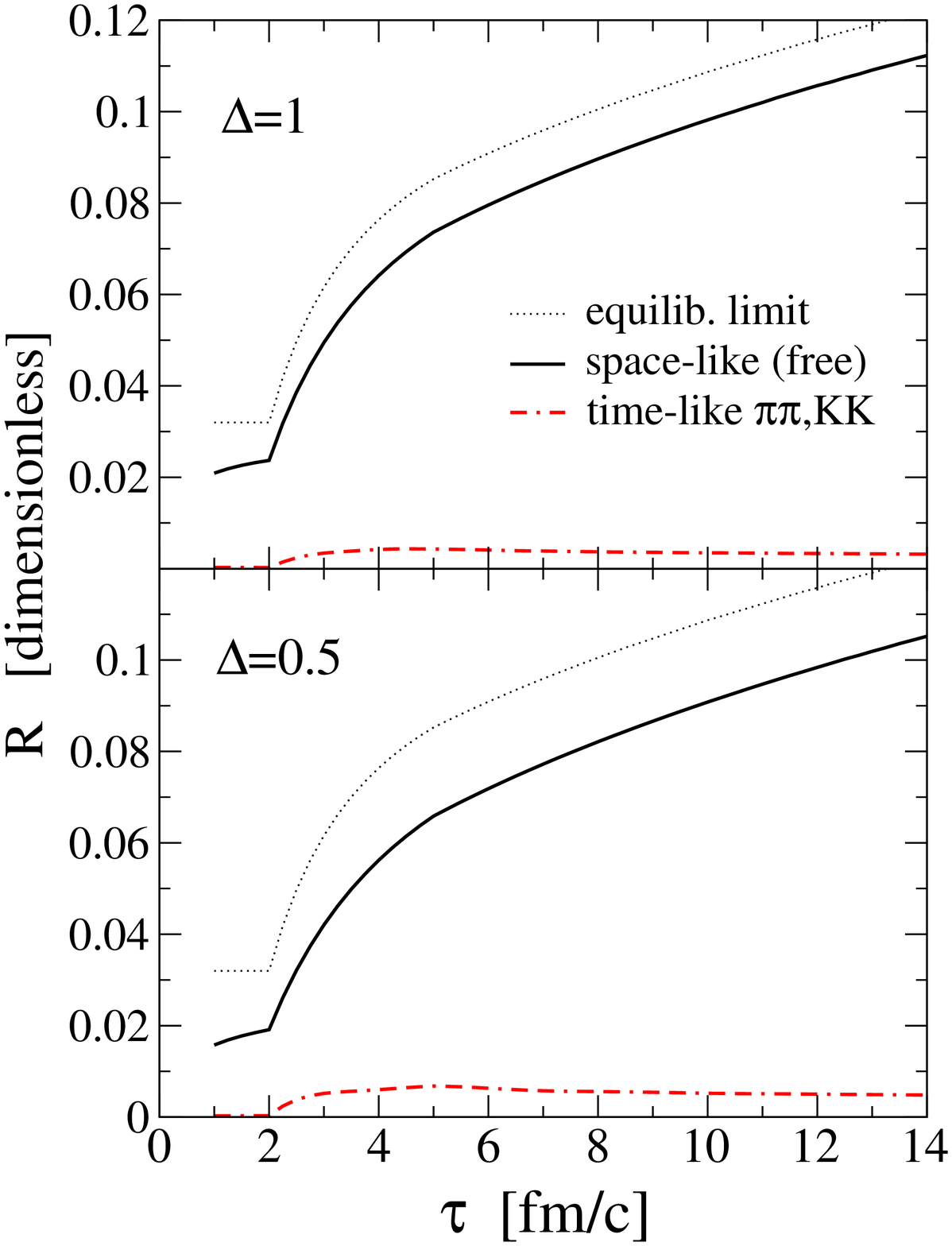,width=8.5cm}
\ece
\caption{Same as Fig.~\ref{fig_dQ2} but with the following
improvements: (i) In the hadronic phase, the entropy density has been
calculated for a hadronic resonance gas, (ii) The e.m.~correlator
includes contributions from $K^+K^-$ states in addition to
$\pi^+\pi^-$ states (both are calculated in the massless limit); and
(iii) In the QGP, massless strange quarks have been included in both
the entropy density and e.m.~correlator.}
\label{fig_dQ2s}
\efg

\subsection{Relation to Other Works}
\label{others}

Our main result from the previous section, Eq.~(\ref{Q2}), bears some
resemblance to the results obtained in Ref.~\cite{SS01}, in particular
to Eq.~(18) therein, which, in the notation of Ref.~\cite{SS01}, reads
\ba
\langle (\Delta Q)^2 \rangle &=& \int \frac{dk}{2\pi} \ 
\frac{\sin^2(k\Delta y/2)}{(k/2)^2} \ \chi_k 
\nonumber\\
 & & \quad \times [1-\exp\lbrace -k^2(\Delta y_{\rm diff})^2\rbrace ] \ .
\label{dQSS}
\ea 
This relation describes the relaxation of the moments of charge
fluctuations in one-dimensional rapidity space through the purely
space-like density-density correlation function $\chi_k$ and the mean
diffusion, $\Delta y_{\rm diff}$, of a particle during time $\tau$.
The $(\sin x/x)^2$ factor above is equivalent to a similar factor in
Eq.~(\ref{Q2}) and arises as a consequence of boost invariant
longitudinal expansion without transverse flow.  The time-dependent
exponential factor in Eq.~(\ref{dQSS}) results from non-equilibrium
effects, which were treated using a Gaussian solution of a
Fokker-Planck equation~\cite{SS01}.  Non-equilibrium effects are not
accounted for in Eq.~(\ref{Q2}) (see, however, our discussion of such
effects in the following section).  Our derivation of Eq.~(\ref{Q2}),
on the other hand, includes the full 4-dimensionally transverse
structure of the conserved isospin current operator, evaluated within
a locally equilibrated relativistic fluid.  This induces additional
quantum effects associated with the finite time scale of the expansion
thereby involving the time-like region of the correlation function.
These features are not captured by the treatment of Ref.~\cite{SS01}, 
in arriving at Eq.~(\ref{dQSS}).

It is worthwhile to point out that, during the dense stages of
evolution in a heavy-ion collision, the assumption of local thermal
equilibrium among the constituents of matter has been gainfully
employed to understand photon and dilepton emission. Calculations in
the CERN SPS regime have shown that a fair agreement exists between
theory and experiment~\cite{RW99,SZ,PP}.  In the context of charge
fluctuations the equilibrium assumption
is supported by the estimates of $\Delta y_{\rm diff}$
presented in Ref.~\cite{AHM00}, and also by the more quantitative
analysis of Ref.~\cite{SS01}, where the diffusion equation
(\ref{dQSS}) was solved with inputs from microscopic transport
calculations~\cite{SS01}.  It was shown there that, due to frequent
rescattering in the hadronic phase of a heavy-ion collision, the
diffusion of charge in rapidity space is effective in keeping
fluctuations close to its equilibrium value (within $\sim$~20\%) until
freeze-out at both SPS and RHIC~\cite{SS01}.  This is consistent with
preliminary experimental data, which point to a dilute hadron gas
in the vicinity of thermal freeze-out.

Notwithstanding these observations, we study in the following section
a somewhat different source of non-equilibrium effects, namely,
fluctuations in the temperature.  This was suggested some years ago as
a means of extracting the heat capacity of the hadronic gas close to
freeze-out~\cite{TF}.  Here we explore connections between
temperature and charge fluctuations, and examine the
consequences of the presence of non-equilibrium effects in heavy-ion
collisions.

\section{DEVIATIONS FROM LOCAL THERMAL EQUILIBRIUM }  
\label{sec_offeq}
Non-thermal fluctuations are generally caused by complex microscopic
and dissipative phenomena.  A microscopic or QCD-based description of
such phenomena is beyond the scope of this paper.  Instead, we use a
macroscopic approach that is based on a stochastic or maximum noise
assumption.  We adduce arguments to show that a non-thermal or
early-stage isospin fluctuation can survive the expansion only if it
is accompanied by a large temperature fluctuation at freeze-out.
Since temperature fluctuations reflect the heat capacity of the
underlying phase, we consider two different possibilities by treating
non-equilibrium effects stochastically through additive (Gaussian) or 
multiplicative (power-law) heat flow.

\subsection{Gaussian Fluctuations}
Let $T(\tau)$ be the temperature at a given proper time $\tau$, and
$T_*=1/\beta_*$ the equilibrium (mean) temperature. Let $T(\tau)$
relax stochastically due to heat flow.  In a phase in which the heat
capacity $C$ depends strongly on $T_*$, thermodynamics implies that
the variance of the temperature distribution is bounded. The
stochastic evolution is ``Brownian'', and the relaxation of $T(\tau)$ is
governed through additive white noise.  The corresponding Langevin
equation is of the form
\be
\frac {dT}{d\tau} + 
\frac 1{\tau_*} \,\left(T(\tau ) -\xi(\tau)  -T_* \right) =0\,\,,
\label{Lang1}
\ee
where $\tau_*$
is the temperature relaxation time, and $\xi (\tau)$ is a white noise, \ie, 
uncorrelated in time and with zero mean. In this case, 
\be
\langle\langle\xi (\tau) \, \xi (\tau')\rangle\rangle= 2D\, 
\delta (\tau-\tau')\,\,.
\label{H2}
\ee
The average in Eq.~(\ref{H2}) is performed over stochastic noise distributions
$\xi$, the strength of which is characterized by the variance $D$. 
The distribution of temperatures 
${\bf P}(\tau, T)$ associated with the additive  Langevin 
equation (\ref{Lang1})  obeys a Fokker-Planck equation~\cite{FP}. Specifically,
\ba
\frac {\partial {\bf P}}{\partial \tau} (\tau, T) = &&
-\frac \partial{\partial T} \left( \frac{(T_* -T)}{\tau_*}\,\,
{\bf P}(\tau, T)\right)
\nonumber\\
&& +\frac 12 \,\frac{\partial^2}{\partial T^2}\,\left( \frac {2D}{\tau_*^2}
\,\,{\bf P} (\tau, T) \right)
\label{H3}
\ea
with a drift rate $\nu_T=(T_*-T)/\tau_*$ and a constant
diffusion rate $\nu_{\rm diff}=2D/{\tau_*^2}$. The general solution is
an evolving Gaussian
\be
{\bf P}(\tau, T) = \sqrt{\frac{1}{2\pi\,\sigma^2(\tau)}} \ 
 {\rm e}^{-(T-\langle T(\tau )\rangle)^2/2\sigma^2 (\tau)}\,\,
\label{PTadd}
\ee
with a stochastic mean
\be
\langle T(\tau ) -T_* \rangle =
 (T(0)-T_*) \ {\rm e}^{-\tau/\tau_*} \,.
\label{H3y}
\ee
Above, single angular 
brackets denote an average with respect to the temperature distribution.
The variance is given by
\be
\sigma^2 (\tau ) = \frac{D}{\tau_*} + 
\left( \sigma^2 (0) - \frac{D}{\tau_*} \right)  
 \, {\rm e}^{-2\tau/\tau_*} \ , 
\label{H3z}
\ee
which reduces to $\sigma^2(\tau\to \infty)=D/\tau_*$ in the stationary case, 
reflecting the thermodynamic limit. It can thus be related to the heat capacity,
$C=T_*^2/\sigma^2$, which depends strongly on $T_*$. 
The stationary temperature distribution takes the form 
\be
{\bf P}_S (T) = {\bf P} (\infty, T) =
\sqrt{\frac C{2\pi\,T^2_*}} \ {\rm e}^{-C (T-T_*)^2/2T_*^2} \ . 
\label{H4}
\ee
The standard Gaussian nature of the fluctuations attests to 
the additive character of white noise.  Obviously, a large heat
capacity strongly limits the character of fluctuations through
exponentially small tails in the stationary distribution,
Eq.~(\ref{H4}).

\subsection{Power-Law Fluctuations}
Here we explore the consequences of a more speculative scenario, that
is, a phase in which the heat capacity $C$ is nearly constant so that
the equilibrium variance grows with the mean temperature squared,
$\sigma^2=T^2_*/{C}$. Such a behavior can be constructed by starting
from a Langevin equation in which the noise is introduced
multiplicatively using the Ito-prescription \cite{BO}. In our case,
this is realized by setting
\be
\frac {dT}{d\tau} + \frac 1{\tau_*} \,
\left[(1+\xi (\tau ))\,T (\tau)-T_*\right] =0 \ , 
\label{H5}
\ee
where, as before, $\xi (\tau)$ is a white noise with zero mean and
variance $2D$. Arguments similar to the ones used in the previous
section yield a Fokker-Planck equation analogous to Eq.~(\ref{H3})
with the same drift rate, $\nu_T=(T_*-T)/\tau_*$, but a non-constant
diffusion rate, $\nu_{\rm diff}=2DT^2/{\tau_*}^2$, which is the key
difference from the result in Eq.~(\ref{Lang1}).  In the steady state,
the distribution of temperatures is now given by
\be
{\bf P}_S (T) = \frac{(1+C)^{1+C}}{\Gamma (1+C)}\,
\left(\frac{T_*}T\right)^{3+C}\,\,\frac {{\rm e}^{-(1+C)\,T_*/T}}{T_*} \,,
\label{H6}
\ee
where the corresponding variance, $\sigma^2=T_*^2/(\tau_*/D-1)$,
allows us to replace $\tau_*/D$ by $1+C$. Note that this is  the
desired behavior for a constant heat capacity.  Larger
temperatures are only power-law suppressed as a direct consequence of
the multiplicative character of the noise. If we further assume that
the temperature relaxation time $\tau_*$ is small in comparison to the
underlying flow time in the evolution, we may average the
temperature fluctuations around the instantaneous mean $T_*=1/\beta_*$
using Eq.~(\ref{H6}). For the leading temperature contribution to the
electromagnetic correlator, \eg, Eq.~(\ref{W1Fab}), this amounts to
\ba
\left\langle f^B(\omega;T)\right\rangle &=& 
\left\langle\frac 1{{\rm e}^{\omega/ T}-1} \right\rangle 
\nonumber\\
 &=& \sum_{n=1}^\infty
\left( \frac 1{1+ n\,\frac{\omega}{T_*\,(1+C)}}\right)^{(2+C)} \ . 
\label{H7}
\ea 
Again, we see that temperature fluctuations turn the exponential
suppression of high energies (implicit in
equilibrium Bose and Fermi distribution functions) into a power-law
suppression. The larger the heat capacity $C$ of the phase considered,
be it hadronic or some sort of mixed phase, the stronger will
be the power-law suppression, and hence the closer the temperature
distribution to a Gaussian, as expected. These results can be readily
adapted to a dynamical description, \eg, a hydrodynamical
simulation, which will be reported in future work.

\section{SUMMARY AND OUTLOOK}                            
\label{sec_concl}

Our purpose in this paper has been to highlight the interdependence  
of fluctuations in isospin charge and electromagnetic
emissivities in the strongly interacting matter encountered during the
dense stages of a heavy-ion collision.
We have established such connections 
both in an infinite system at rest, as well as in a
relativistically expanding fluid to mimic some of the dynamics present
in high-energy nuclear collisions.

In the static case, we have emphasized that well-known relations
between the fluctuations of the charged constituents of matter and its
electric screening mass imply an intimate connection to the emission
of dileptons/photons.  Both of these can be calculated from the
retarded current-current correlation function, or the photon
self-energy, in hot and dense partonic or hadronic matter.
Consequently, analyses of dilepton and photon observables that are based
on the e.m. correlation function can be systematically applied to
study medium effects on fluctuations during 
the dense stages of a heavy-ion collision.  
Two hadronic approaches were used to exemplify that close
to the expected phase boundary, hot and dense hadronic matter might
exhibit fluctuations of similar magnitude as an equilibrated partonic
medium (QGP).  Clearly, further work is required to improve these
estimates.
  
In the case of an expanding fluid, we derived an expression for the
fluctuations in a locally equilibrated fluid element as a function of
proper time and over a finite range in pseudo-rapidity.  Isospin
conservation, which governs the 4-dimensionally transverse
Lorentz-structure of the correlation function, implies that
fluctuations depend on both transverse and longitudinal components.
Furthermore, the finite time scale $\sim~\tau_{fo}$ dictated by the
strong interaction dynamics in a nuclear collision implies that both
space-like and time-like processes are involved.  At a given
freeze-out proper time $\tau_{fo}$ and within a rapidity window of
width $\Delta$, the thermal isospin fluctuations are dominated by
constituents with typical frequencies $f\approx 2\pi/\tau_{fo}$ and
wavelengths $\lambda\approx \pi\tau_{fo}\,{\rm sh}(\Delta/2)$.

Our analysis of non-equilibrium effects associated with possible
temperature fluctuations indicates that a large isospin
fluctuation at the early partonic stage can survive the course of a
relativistic heavy-ion collision only if a large temperature
fluctuation occurs simultaneously at freeze-out in the same event.
Given a large and strongly temperature-dependent heat capacity of a
hadronic gas, this scenario seems unlikely.

Prospects for the future include: (1) A determination of the Debye
mass of the system at the dilute freeze-out stage of a heavy-ion
collision; this could be accomplished solely through measured results
of particle multiplicities and particle correlations using
Hanbury-Brown and Twiss analyses.  An intriguing question in this
connection is whether or not the Debye mass at freeze-out remains
constant from AGS through LHC energies; and (2) A delineation of the
role of charge fluctuations during the dense stages of the collision
through a study of electromagnetic emissivities; this is a more
challenging task inasmuch as an understanding of the measured dilepton
and photon yields will necessarily require a detailed theoretical
analysis to unravel the phase structure of the rapidly evolving
produced matter.

Our ongoing investigations include a dynamical simulation of the
collision, in which the effects of strong interactions are studied
through the equation of state and the effects of fluctuations in dense
matter are studied through its electromagnetic radiation.

\vskip .4cm
\section*{\bf ACKNOWLEDGEMENTS}
\vskip .2cm
We thank Edward Shuryak for helpful discussions.  This work was
supported in part by the US DOE grant DE-FG02-88ER40388 and by the
BMBF.

\end{document}